\documentclass[]{article}
\input psfig.sty

\begin{document}
\begin{center}
\Large SIR JOHN HERSCHEL AND THE STABILITY OF SATURN'S RING

\vspace{1.0cm}

\normalsize
{\it by Alan B. Whiting \\
University of Birmingham}
\end{center}

\vspace{1.0cm}

\large
In a pioneering exposition of mathematical astronomy for the public,
Sir John Herschel attributed the stability of the ring of Saturn to
its being eccentric with respect to the planet and lopsided
(asymmetric in mass) by a minute amount.  Tracing the sources and
effects of this error reveals several lessons of general relevance
to science: on the formulation and interpretation of calculations,
the use of cutting-edge observations and the combining of observations
with theory.  I emphasise the phenomenon of reinforcing errors.
\normalsize

\vspace{1.0cm}

\noindent {\it Astronomy for the Public in 1833}

Sir John Herschel (1792-1871) was not only a major figure in astronomy and
related sciences in the middle of the nineteenth century, he was
a pioneer in explaining the workings of a highly mathematical subject to
the general public.  His {\it Treatise on Astronomy}$^1$ of 1833 wrestled
with the difficulty of describing the workings and results of
mathematical physics (including the abstruse theory of the mutual
perturbations of the planets) without actually displaying the equations.
He must have been at least moderately successful, since the
{\it Treatise}, in its somewhat rewritten form as {\it Outlines of
Astronomy}$^2$, went to eleven editions, with the latest coming out in 1869
(and being reprinted at least as late as 1902).

Perhaps surprisingly, almost all of his 1833 book could be used in a 
university course in astronomy today.  The knowledge set out in it has
stood the test of time (especially the details of the planetary orbits)
and Sir John was careful in warning his readers about
what was speculation and what was simply unknown.  (Of course, a book
written before the discovery of Neptune, the use of spectroscopy and
the formulation of thermodynamics has its limitations for later use,
a point made as early as the 1902 reprinting).

However, for this study I am going to examine one of the mistakes, one
of the few statements made by Herschel without warnings or caveats that
turned out to be wrong.  The episode illuminates the workings of our
science, some details of its progress and a bit of the philosophy of its
practicioners.

\vspace{0.5cm}
\noindent{\it Herschel on Saturn's Ring}

After introducing Saturn, describing the appearance of the planet and giving
various numbers (dimensions, distance and such), Herschel$^1$ says
\begin{quote}
Although the rings are, as we have said, very nearly concentric with the body
of Saturn, yet recent micrometrical measurements of extreme delicacy have 
demonstrated that the coincidence is not mathematically exact, but that
the center of gravity of the rings oscillates round that of the body describing
a very minute orbit, probably under laws of much complexity.  Trifling
as this remark may appear, it is of the utmost importance to the stability of
the system of the rings. Supposing them mathematically perfect in their
circular form, and exactly concentric with the planet, it is demonstrable
that they would form (in spite of their centrifugal force) a system in a
state of {\it unstable equilibrium}, which the slightest external power
would subvert---not by causing a rupture in the substance of the rings---but
by precipitating them, {\it unbroken}, on the surface of the 
planet. (\S 444, p.\ 284; here, as elsewhere, I retain the emphasis of the
original publication)
\end{quote}
In the same section but
on the next page he goes on to say that, to ensure stability,
\begin{quote}
\dots it has been shown this it is sufficient to admit the rings to be
{\it loaded} in some part of their circumference, either by some minute
inequality of thickness, or by some portions being denser than others.
Such a load would give to the whole ring to which it was attached somewhat
of the character of a heavy and sluggish satellite \dots
\end{quote}
There are a number of things to take issue with in this section and
neighboring ones.
Among them, Herschel later confusingly insists
that `we have no proof' of the ring being loaded, and offers some comments
on stability
that are more difficult to interpret.  For our purposes,
however, there are four important points: the assumption that the ring is
a solid body; the assertion that a uniform
ring would be unstable; that the smallest amount of non-uniform mass
distribution would impart stability; and that an eccentricity, a displacement
of the rings from perfect concentricity with the planet, had been
observed.  The idea of the rings being solid would lead to a lengthy
historical investigation that I will not attempt here; I note it simply as
a background to the issue of stability.  The second two assertions belong
to dynamical theory and, of course, could not be expounded in a work designed
without equations; we will examine their source in the next section.
I will look at the observational story at in the section following. 

\vspace{0.5cm}
\noindent{\it The theory of rings}

Not only is a popular work without mathematical derivations, it generally
lacks the
footnotes that would direct one quickly and easily to their origin; so
it is in this case.
A secondary source,
Alexander$^3$ describes a stability calculation which sounds much like
Herschel's assertions, attributing it to a memoir by Laplace$^4$.
However, a reading of that memoir (very helpfully published on line
by the Biblioth\'{e}que National de France) reveals nothing of the kind.
Alexander appears to be paraphrasing a description by Proctor (no
specific reference is given), so perhaps he mixed up his references or
Proctor did; in either case we appear to have another instance of the
Law of Propagation of Bad Data.

However, both Herschel's assertions and Alexander's description fit
very well a section of Laplace's {\it M\'{e}canique
C\'{e}leste}$^5$.  Conveniently for many of us it was translated into
English and provided with an extensive commentary
(roughly doubling its size) by Nathaniel Bowditch, an American mathematician
and navigator, his first volume appearing in 1829.  This is the version
I have consulted, though (considering publication dates and the slowness
of trans-Atlantic travel at the time) Herschel more probably referred to the
original French publication of 1799.

In Book III,
Chapter VI, \S 46, Laplace considers the stability of a uniform solid hoop
under the gravitational pull of a planet.  If the hoop is centred on the planet
it is in equilibrium, since each part of the hoop is subject to a force that
is exactly counteracted by a part on the opposite side of the planet.  If the
hoop is disturbed, what happens?

Setting up the integral for gravitational potential to decide the
matter, Laplace found no simple form but
succeeded in transforming it into a series that is a monotonically 
decreasing (increasingly negative)
function of the displacement of the hoop from concentricity.  In his words,

\begin{quote}
\ldots therefore {\em the centre of Saturn repels the centre of this circular
homogeneous ring; and whatever be the relative motion of the second centre
about the first, the curve it describes, by this motion, is convex towards
Saturn; the centre of this circular ring must therefore recede more and
more from the centre of the planet, until its circumference shall finally
come in contact with the surface of the planet.}

\ldots

{\em Hence it follows, that the separate rings which surround the body of
Saturn, are irregular solids, of unequal widths in the different parts of
their circumferences; so that their centres of gravity do not coincide
with their centres of figure.  These centres of gravity may be considered as
so many satellites, which move about the centre of Saturn, at distances
depending on the inequalities of the parts of each ring, and with velocities
of rotation equal to those of their respective rings.} (pp.\ 515-516) 
\end{quote}

In his extensive footnote
Bowditch explains the second paragraph by showing that {\em for a geometrically
centred ring} (a tacit, but important, assumption) the centre of gravity
of the hoop is moved
off the geometric centre by its mass loading, and that the centre of gravity
is now pulled toward the centre of the planet rather than being expelled.

I will take each assertion in turn.  First,
there is one major flaw in Laplace's proof of the instability of a symmetrical
ring: the failure to take motion into account.  Although the rotation of the
rings was clearly known to him (and indeed used in calculations in the previous 
section) it is not included in the mathematics of stability.  Perhaps the
fact that the planet can exert no net torque on a symmetric ring led him
to dismiss it as a factor.  But even in a symmetric case, as a spinning top
or a figure skater, the rotation does have an effect on the dynamics
through possible changes in the moment of inertia.  A satellite, indeed,
would fall directly into the planet without its motion of revolution.

The mention of a satellite brings up another problem: the confusion of
static with dynamic stability; or, perhaps, the failure to notice that
they are different concepts.  This I will take up in some detail below.

In the proof of the stability of the loaded ring there are two errors, either
of which reduces it to nonsense.  The first is that a loaded ring that is
geometrically centred on the planet is {\em not} in equilibrium---obviously,
since the forces are not in balance; so one {\em cannot} take this position as
the beginning of a stability analysis.  Second, the motion of the centre of
mass is quite secondary to that of the ring itself.  The loaded ring is still
drawn towards the planet along one part of its circumference; and Laplace is
silent on what might happen after the centres of gravity of the loaded ring
and the planet coincide.  We are left, at the very best, with a feeling that
a loaded ring might behave something like a satellite of the planet, and
a feeling or analogy is not a proof.

So the dynamical basis of Herschel's assertions, the instability of a
symmetrical ring and the stability of a loaded one, has been shown to be
mistaken.  (Note that neither assertion has been proven to be wrong!  I
will consider both questions in detail below.)  All in all, this
is a disappointing performance by three very competent mathematicians.

\vspace{0.5cm}
\noindent{\it Measuring the rings}

An account of
the beginning of the observational side of this episode is
to be found in the {\it Philosophical
Magazine} for 1828 July$^6$.  On pages 62-3 are found summaries of two
letters.  First, one from Professor Harding of G\"{o}ttingen
related how Heinrich Schwabe had gained the impression that
the eastern side of the rings of Saturn were farther from the planet than
the western side, and the two had agreed over several months (1827 December
to 1828 May) that it appeared to be so.  Harding thought it was an
`optical deception' but could not explain it, and asked astronomers with
better telescopes and more precise instruments to look into the matter.

The response of James South to this request appears immediately afterward in the
same issue.  South and several other observers, including Sir John Herschel,
observed the planet on 1828 April 26, 29 and May 8 using South's
`five-foot' refractor (the lens of which would have been about five
inches in diameter).  The average of 35 micrometer measurements made by
South and Herschel gave the space between the planet and the ring as
3.535'' on the west side and 3.607'' on the east, a difference South
considered insignificant.  On April 26 each of the two observers
made ten measurements, Herschel getting an average of 3.612'' on the west,
3.442'' on the east, while South's averages were 3.331'' on the west and
3.502'' on the east.  Still, Herschel thought the gap between the ring
and the planet appeared larger on the east  side, and six of the seven
observers present agreed.  However, South writes that Saturn was low in
the sky and the observations were `far from satisfactory.'

The challenge was next taken up by F. G. W. Struve, the consummate observer of
double stars and perhaps the best person in the world at the time at
measuring tiny features on the sky.  His report appears in 1828 May$^7$.
Using the 9.5'' Fraunhofer refractor at Dorpat (now Tartu) he observed
Saturn on several nights in 1828 March and April.  His results, given
in seconds of arc between the planet and the outer edge of the
rings and corrected
for the changing distance from the Earth
to the planet and the phase angle, are as
follows:

\begin{tabular}{lcrrr} \\
Date (1828) & Number of & West side & East side & difference \\
 & observations & & & \\
March 29 & 1 & 11.272 & 11.390 & +0.118 \\
April 7 & 2 & 10.996 & 11.250 & +0.254 \\
April 7 & 3 & 11.148 & 11.260 & +0.112 \\
April 9 & 4 & 10.931 & 11.243 & +0.312 \\
April 10 & 2 & 11.238 & 11.485 & +0.247 \\
April 21 & 3 & 11.060 & 11.238 & +0.178 \\
total/avg. & 15 & 11.073 & 11.288 & +0.215 \\
\end{tabular}

\begin{figure}
\centerline{\psfig{file=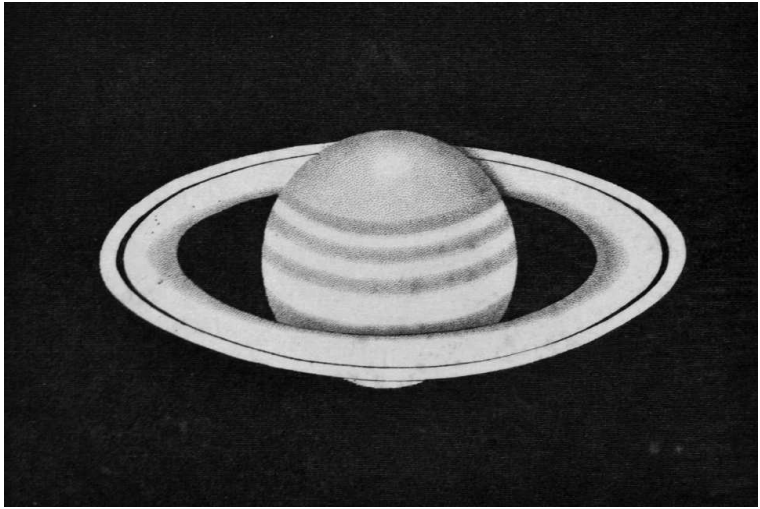,width=10.0cm}}
\caption{An engraving of Saturn from Sir John Herschel's {\it Treatise
on Astronomy} from 1833$^2$.  The eccentricity as measured by F. G. W.
Struve amounts to slightly less than half a millimetre on this figure as
printed.}
\label{saturn}
\end{figure}

Struve gives the likely error for a single observation of 0.095'', which is
consistent with the table, and thus the likely error for the average of
fifteen measurements is 0.024.''  Before his observations Stuve `considered
the difference \ldots to be an optical illusion,' but afterward thought
that his measurements `prove the excentricity of the sphere within its 
rings in the most conclusive manner.'

Before we consider the observations in detail it is worth asking
whether it is at all reasonable that Struve could distinguish such
tiny details in the first place.
Two-tenths of an arc
second is a fraction of the Airy disk of the telescopes used;
it is not trivial to
see or measure this amount.  But based on personal experience
I think it is possible. 
Using the eight-inch Alvan Clark refractor at the U.\ S.\ Naval Academy,
of the same type as Struve had at his disposal though slightly smaller
and newer, on a particularly good night I 
noticed (without looking for it) that the Galilean moons of Jupiter were
all of different sizes.  This means I was distinguishing between four
disks of 0.8'' to 1.4'' diameter, differences of one or two tenths of an
arc second, just what we are considering.  The observations are of a
different kind and I made no measurements, but I am still led to
conclude that Struve's report is {\em a priori} plausible.

There are, however, difficulties when we consider the efforts of all the
observers.  If the eccentricity was so clear to most observers in
the party of Herschel and South, why couldn't they measure it?  Indeed,
Herschel's own figures on 26 April
went the wrong way.  And if the conditions were
so poor that measurements of one or two tenths of an arc second could not
be made reliably, it strongly suggests that whatever was giving the
appearance of eccentricity was not an actual displacement of the rings
of this amount.

A hint that Struve's observations in themselves were not quite what they 
were taken
to be could be gained from looking hard at the table he gave.  If we consider
each line to be a separate observation instead of an average of a few
(which is not quite the correct way to look at them, but will serve to
prove this point)
we find the standard deviation of the western measurements to be 0.128
and of the eastern 0.097.  If they were statistically independent we would
expect the standard deviation of
their difference to be these added in quadrature, 0.161; instead
we have half that, 0.081---indeed, {\em less} than that of either
measurement alone.  There is thus some reason to suspect that the
standard statistical methods as used by Struve
may not be applicable to this set of
figures; at the very least a careful review from the beginning needs
to be made.

Thus there were reasons to be suspicious of the reported observation of
eccentricity in Saturn's rings.  There is, however, a rather stronger
reason to question them, one that does not seem to have been considered
at the time: a dynamical one.  All astronomers involved were well
aware that Saturn's ring revolves quickly around the planet, 
taking about ten and a half hours for a revolution.
Why, then, should the eccentricity, especially if tied to a particular
point on the ring (where the mass-load is located), stay on the same side?
It apparently did so not only over days but over months and on both sides of
opposition.  While the others may have been only pursuing a puzzling
observation on the edge of current capabilities, Herschel explicitly
put observation and theory together and should indeed have noticed the
dynamical problem.

\vspace{0.5cm}
\noindent{\it Afterward: Maxwell and another Struve}

Herschel's eccentric Saturn seems to have had little effect on 
the progress of astronomy.  Although he retains essentially the same
passage in his later popular astronomy books (even in the edition of
1869), I have found few other references to the matter.

As far as observation goes, the elder Struve's result of 1828 was not
confirmed.  His son Otto Struve, using the much larger refractor at
Pulkovo in 1851, found no significant eccentricity$^8$\footnote{In that
publication he did, however, assert that the ring system as a whole 
was of a different size than earlier measurements allowed, and thus
must be changing on the time scale of decades to centuries.  To follow this 
observational tangent would obviously take me too far from Herschel
and the matter at hand.  Of relevance to our question is Otto Struve's
comment that his measurements of the inner edge of the rings are
of less precision because of the difficulty of defining the border
of the newly-discovered crepe ring}.  What caused the strong impression
of eccentricity, and what the elder Struve actually measured, remain
unknown.

The dynamical theory of Saturn's rings received more attention.  What
came to be considered the definitive answer was
set out by James Clerk Maxwell at the outset of his scientific career$^9$.
The Adams Prize for 1856 set the problem of calculating the
dynamics of the ring system of Saturn, allowing the contestants to
choose what form they were to assume the rings took.  That is, 
`It may be supposed that
(1) they are rigid; (2) that they are fluid and in part aeriform;
(3) that they consist of masses of matter not materially coherent'
(p. 286).  Maxwell looked at each possibility in turn, but it is only
the first that concerns us here.

Maxwell begins by pointing out that a solid object of planetary size,
especially one so broad and thin, cannot maintain itself by its own
strength `though it were made of the most rigid material known on
Earth' (p. 287), a point also made by both Laplace and Herschel,
and thus that it must be in `dynamical equilibrium'
(p. 286), a very important phrase.  Considering Laplace's results,
Maxwell says that 

\begin{quote}
he proves most distinctly (Liv. iii Chap. vi) that
a solid uniform ring cannot possibly revolve around a central body in
a permanent manner, for the slightest displacement of the center
of the ring would originate a motion that would never be checked, and 
would inevitably precipitate the ring upong the planet, not
necessarily by breaking the ring, but by the inside of the ring falling
on the equator of the planet (pp.\ 293-4)
\end{quote}
However,
\begin{quote}
I have not discovered, either in the works of Laplace or in those of
more recent mathematicians, any investigation of the motion of a ring
either not uniform or not solid (p.\ 294).
\end{quote}

So Maxwell accepts Laplace's result on instability, to the point of
repeating the claim that the ring would be unbroken upon striking the
planet's surface.  (This is interesting in the context of his emphasis
on the role rigidity cannot play in the whole situation.  It appears
that all of our scientists, from Laplace on, have ignored the very
serious differential stresses that would act on the ring during its fall
onto the planet.)
However, Laplace's assertion of the stability of a loaded ring is simply
ignored.  Perhaps Maxwell thought it merely an analogy or plausibility
argument (as indeed it is), rather than a serious attempt at calculation.

In using the phrase `dynamical equilibrium' above Maxwell has hinted at
a deeper and more complex idea of stability than that exhibited by
Laplace, Bowditch or Herschel.  Before actually starting his work
he makes this explicit:

\begin{quote}
There is a very general and important problem in dynamics, the solution of
which would contain all the results of this essay and a great deal
more.  It is this---

`Having found a particular solution of the equations of motion of any
material system, to determine whether a slight disturbance of the motion
indicated by the solution would cause a small periodic variation, or
a total derangement of the motion.' (pp.\ 295-6)
\end{quote}

This is the criterion of dynamical stability that Maxwell uses, though
he does not use the term explicitly here.

Maxwell then addresses Saturn's rings as a collection of thin, rigid hoops,
circular but not necessarily uniform in mass around their circumference,
and investigates the dynamical stability of one representative example.
He writes down the equations of motion, allowing for the displacement
of the centre of mass of the ring from that of the planet; rotation of
the ring about its own centre of mass; and rotation of the line joining
the centres of mass.

He next imposes, as his dynamic equilibrium, a state of uniform rotation.
This is interpreted to mean that the `position of the centre of the
planet with respect to the ring does not change' (p.\ 299) and that the
rotation of line joining the centres of mass proceeds at a constant
rate.  As a condition for this state Maxwell finds that the gravitational
potential of the ring must be at a stationary point; in other words, the
planet must occupy a (possibly local) minimum or maximum of the hoop's
potential field\footnote{Or, strictly speaking, an inflection point; but 
it's hard to see how this could actually happen for any reasonable hoop.}.
This means that Maxwell's treatment is not general enough to include,
as a limiting case, all the ring's mass concentrated at a point (that is,
a small satellite)\footnote{Indeed, it raises some question about the
general applicability of Maxwell's treatment.  Dealing with such a question
in detail, however, would take us rather far from Herschel and Laplace.
In any case, Maxwell's further sections
on the stability of other kinds of ring seems to take care of any doubts
we might have on the subject.}.

These appear, at first sight, to be a quite restrictive set
of conditions.  Certainly one could imagine a hoop in some sort of
motion in which distances and angular velocities vary within limits;
that is,
one might allow the analogy of a Keplerian ellipse, rather than requiring
the equivalent of a strict circular orbit.  Maxwell does not discuss his
reasoning.  It can be justified, however, by appealing to the structural
strength of the hoop: if the motion is not uniform, the centrifugal force
cannot be expected to balance gravitational force exactly
everywhere, and stresses
would be set up in the hoop that no material could support.

Maxwell then proceeds to allow small variations in the angular quantities
and distance, puts them in exponential form, and investigates the
conditions under which the exponents remain imaginary or negative: 
a perturbation analysis of the familiar kind.  Expanding the mass
distribution around the ring in a Fourier series, he finds that only the
first few coefficients have any effect on the (infinitesimal) stability,
allowing a useful level of simplification.

The salient result, for Maxwell, was that for a loaded ring, an
object equivalent to `a single heavy particle placed at a point on
the circumference of the ring' (p.\ 310) at this level of Fourier
approximation, the ring must be off centre by an amount between
0.81565 and 0.8279 of the radius in order to be stable.  This 
not only requires a very
`finely-tuned' system (to use a current term), it is quite ruled out
by observation.

Maxwell then proceeds to investigate the other possiblities for the
nature of Saturn's rings, eventually concluding that they could only
be made up of innumerable tiny objects, each in a separate orbit 
around the planet.  This is the result most remembered from his
analysis and the one accepted today.
 
\vspace{0.5cm}
\noindent{\it Another look at stability}

Having outlined the results of Laplace and Maxwell on the subject of
Saturn's rings considered as rigid hoops, it is instructive to look at
another approach.  This uses energy and angular-momentum arguments,
a more modern way of working, though probably understandable by these
consummate mathematicians.

\begin{figure}
\centerline{\psfig{file=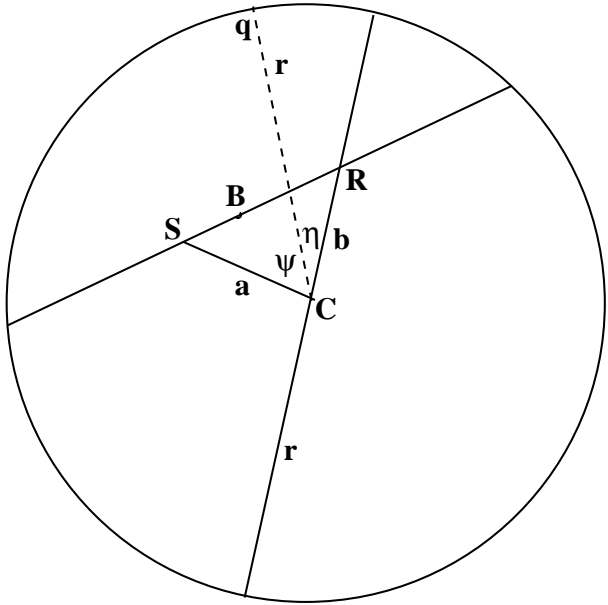,width=8.0cm}}
\caption{Figuring the dynamic stability of a solid ring about a planet.
The ring is drawn, with geometric centre at $C$ and radius $r$; if its
mass distribution is not uniform the centre of mass is offset by a
distance $b$ and is located at $R$.  Its rotation about its centre of 
mass is measured by the angle $\phi$ (not shown) from a fixed reference 
direction.  The centre of mass of the planet,
idealised as a sphere, is at $S$, offset from the centre of the ring by
the distance $a$ and from the barycentre of the whole system by the
distance $SB$.  The line joining the centres of mass, $SR$, rotates
as measured by the angle $\theta$ (also not shown).
A typical point on the ring is represented by $q$.}
\label{ringstable}
\end{figure}

The situation is diagrammed in Fig.~\ref{ringstable}.  The planet's centre
of mass is at $S$, that of the ring at $R$, and these rotate around the
barycentre $B$ as measured by
the angle $\theta$ between the line of centres and a fixed direction
(not drawn, for clarity).  The ring rotates about its
centre of mass as measured by an angle $\phi$ between the line $CR$
and a fixed direction (also not drawn) which may be different from
$\theta$ (in particular, it may change at a different rate).  With the
mass of the planet $M$ and of the ring $m$, we may write down the
kinetic energies of angular motion for the planet about the barycentre,
the ring about the barycentre and the ring about its centre of mass 
respectively as
\begin{eqnarray}
T_M &=& \frac{1}{2} M (SB)^2 \dot{\theta}^2 \nonumber \\
T_R &=& \frac{1}{2} m (BR)^2 \dot{\theta}^2 \nonumber \\
T_r &=& \frac{1}{2} m (r^2-b^2) \dot{\phi}^2
\end{eqnarray}
The kinetic energy of linear motion is
\begin{equation}
T_L = \frac{1}{2} M \dot{(SB)}^2 + \frac{1}{2} m \dot{(BR)}^2
\end{equation}
The total angular momentum, which is constant, is
\begin{equation}
J = M (SB)^2 \dot{\theta} + m (BR)^2 \dot{\theta} + m (r^2- b^2) \dot{\phi}
\end{equation}
The distribution of mass about the ring is given by a function $\nu(\eta)$,  
where $\eta$ is the angle measured at $C$ from
the line $CR$.  It is related to the eccentricity of the centre of mass of
the ring by
\begin{equation}
b = \int_{0}^{2\pi} \nu r \cos \eta d\eta
\end{equation}
and, of course, is positive (or zero).  We will also need the angle $\psi$,
measured at $C$ from the line $SC$.

The gravitational potential energy of the ring with respect to the planet
is
\begin{equation}
V = -GMm\int_{0}^{2\pi} \frac{\nu d \psi}
{\sqrt{a^2+r^2-2ra\cos \psi}} 
\end{equation}
(note that, since both $\psi$ and $\eta$ are measured from $C$ and in each 
equation given go around the full circle, we may freely convert between
them).  To work out the dynamics we shall have to shift from $a$ to
$SR$ as the variable of interest; although $a$ is more directly 
observable, $SR$ allows a more direct connection to the expressions
of kinetic energy.  While $a$ and $b$ are small all three quantitites
are of the same order, so in examining the situation close to symmetry
it suffices to work out the behaviour of either $a$ or $SR$.  The
conversion between them is given by
\begin{equation}
a^2 = b^2 + (SR)^2 - 2 b SR \cos (\phi - \theta).
\end{equation}

To clean things up a bit, we introduce the parameters $\mu = m/M$,
$\sigma = SR/r$, $\alpha = a/r$ and $\beta = b/r$.  

The total kinetic energy is
\begin{equation}
T= \frac{M}{2r^2}\frac{\mu}{\mu+1}\dot{\sigma}^2 +
\frac{M}{2r^2}\frac{\mu}{\mu+1}\sigma^2 \dot{\theta}^2 +
\frac{M}{2r^2}\mu \left( 1-\beta^2 \right) \dot{\phi}^2.
\end{equation}
We now introduce the parameter $\gamma = \dot{\phi}/\dot{\theta}$, the
rate at which the ring rotates about its centre of mass compared to the
rate of rotation of the centres of mass about the barycentre; for a
symmetrical ring it will be unity.  We can now write the total angular
momentum as
\begin{equation}
J = \frac{M}{r^2} \left( \frac{\mu}{\mu+1} \sigma^2 + 
\mu (1-\beta^2) \gamma \right) \dot{\theta}
\end{equation}
and use this expression to substitute for $\dot{\theta}$ in the
kinetic energy, giving
\begin{equation}
T=\frac{\left( \frac{M}{2r^2}\frac{\mu}{\mu+1}\sigma^2 + \frac{M\mu}{2r^2}
(1-\beta^2) \gamma^2 \right) J^2}
{\left( \frac{M}{r^2}\frac{\mu}{\mu+1} \sigma^2+\frac{M\mu}{r^2} (1-\beta^2) 
\gamma \right)^2} + \frac{M}{2r^2}\frac{\mu}{\mu+1} \dot{\sigma}^2.
\end{equation}
There is some obvious cleaning up to do in this expression.
In addition, we will want to compare the total angular momentum of the
eccentric ring to the situation in which all the angular momentum
is contained in the ring material performing a circular orbit about
the planet; in the latter case the angular momentum is
\begin{equation}
J_{0}^2 = GMm^2r
\end{equation}
and we will introduce the angular momentum parameter $j=J/J_0$.

We are now in a position to write down a useful expression for total energy
of the system, arriving at
\begin{eqnarray}
E &=& \frac{1}{2} M r^2 \frac{\mu}{\mu+1} \dot{\sigma}^2 \nonumber \\
&+& \frac{GMm}{r}\frac{j^2 \left(\frac{\sigma^2}{\mu+1}
+\left(1-\beta^2\right) \gamma^2 \right)}
{\left(\frac{\sigma^2}{\mu+1}+\left(1-\beta^2\right) \gamma \right)^2}
\nonumber \\
&-& \frac{GMm}{r}\int_{0}^{2\pi} \frac{\nu d \psi}
{\sqrt{1+\alpha^2 - 2 \alpha \cos (\phi - \theta)}}.
\end{eqnarray}
This may be rearranged to form 
\begin{eqnarray}
\frac{1}{2} M r^2 \frac{\mu}{\mu+1} \dot{\sigma}^2 &=& E - \frac{GMm}{r} \times
\nonumber \\
\left\{ \frac{j^2 \left(\frac{\sigma^2}{\mu+1} 
+\left(1-\beta^2\right) \gamma^2 \right)}
{\left(\frac{\sigma^2}{\mu+1}+\left(1-\beta^2\right) \gamma \right)^2} \right.
&-& \left. \int_{0}^{2\pi} \frac{\nu d \psi}
{\sqrt{1+\alpha^2 - 2 \alpha \cos (\phi - \theta)}} \right\}
\label{master}
\end{eqnarray}
the equation 
expressing the dynamics of a particle of 
mass $Mr^2 \mu/(\mu+1)$, (one-dimensional) coordinate $\sigma$ and (constant)
total energy $E$ in an effective potential given by the expression
involving the brackets.  The radical in the denominator of the integral, in
terms of useful quantities, is
\begin{equation}
1+\beta^2+\sigma^2-2\beta \sigma \cos (\phi - \theta)
-2 \cos \psi \sqrt{\beta^2+\sigma^2-2\beta \sigma \cos (\phi-\theta)}.
\end{equation}
For the ring to have stable motion, the effective potential as a function
of $\sigma$ must have a local minimum.

Let us consider first the completely symmetrical situation.  In that case
$\beta = 0$, $j = \gamma = 1$, $\sigma = \alpha$, $\phi = \theta$
and $\nu = 1/2\pi$, and Eq.~\ref{master}
gives
\begin{equation}
V_{\rm eff} \propto \frac{\mu +1}{\sigma^2+\mu+1} -\frac{1}{2 \pi}
\int_{0}^{2\pi} \frac{d \psi}{\sqrt{1+\sigma^2-2\sigma \cos \psi}}.
\end{equation}  
The left-hand term, the one introduced by the rotation of the ring,
is a monotonically decreasing function of $\sigma$, and not a very
strongly varying function at that.  The integral is a monotonically
increasing function of $\sigma$, with increasing derivative, becoming
singular at $\sigma = 1$; thus the whole expression is monotonically
decreasing.  There is no local minimum and motion is unstable.
The addition of rotation does not stabilise the ring, though it does
modify the dynamics somewhat.  Herschel's
second assertion about Saturn's rings is correct.

Consider, now, Eq.~\ref{master} in the case of a ring close to symmetry.
Since $\beta$ is small and always occurs added to or subtracted from
quantities that are not, it will have no qualitative effect on
$V_{\rm eff}$.  Similarly, $j$ and $\gamma$ will be close to unity,
the first only rescaling the angular-momentum term slightly and
the second (since it shows up to the same degree in numerator and
denominator) having no strong effect.  A small eccentricity due to
asymmetric mass does not stabilise the ring.  Herschel's third assertion
concerning Saturn's rings is incorrect.

What, then, of Herschel's intuition that a loaded ring would act like
a `sluggish' satellite?  If we take the limiting case in which all the
mass of the ring is concentrated at a single point, we obtain
\begin{equation}
V_{\rm eff} = \frac{GMm}{r} \left\{ \frac{j^2 (\mu+1)}{\sigma^2}
- \frac{1}{\sigma} \right\}
\end{equation}
which indeed has a minimum\footnote{For a ring/satellite of negligible
mass compared to the planet ($\mu = 0$) and having the same angular momentum
as a symmetrical ring centrally placed ($j=1$) 
the minimum is at $\sigma = 2$, placing
the planet {\it on} the (now merely conceptual) ring.  This suggests
that any stable configuration would also be rather eccentric, in agreement
with Maxwell's result.}.  A `ring' with the proper combination of total
energy and angular momentum would be confined within certain values of
$\sigma$; alternatively, if a circular orbit at the minimum value of
$\sigma$ is perturbed, the ring stays within a small distance of this
value.  (These are {\it different} definitions of stability, as will be
discussed below.)

Somewhere, then, in parameter space between a symmetric ring and a
totally asymmetric one there is a border between stability and instability.
Seeking to define the border in a multidimensional parameter space
would be tedious (it depends on the distribution of mass in the ring,
which can in principle be very complicated).
For our purposes it is sufficient to
note that it is not near the symmetric case, but that Herschel's intuition
is qualitatively correct.

\vspace{0.5cm}
\noindent{\it Cutting-edge observations, dynamic stability and philosophy}

Although all the details of this episode concern questions, techniques
and equipment that have long been left behind by astronomy, there are
a number of lessons that are every bit as applicable for us now as they
were for Herschel and his contemporaries.  I will divide them up into
three areas: the treatment of observations at the forefront of research;
the formulation and interpretation of mathematical analysis; and some
aspects of the practical philosophy of science.

Observationally, we have a situation in which one measurement with
cutting-edge instrumentation (placing all of Struve's numbers together as
one measurement with a given probable error) gives a formally significant
result.  It is not confirmed by other efforts made with admittedly
less capable equipment.  Certainly the same kind of situation has arisen
since, and no doubt will arise again.  One's reaction to such a
singular observation will generally depend on one's prejudices,
perhaps dressed up with Bayesian numbers; but in any case a single
result from a single source is difficult to handle.

In this general situation I suggest that a useful attitude is that of
Sir A. S. Eddington in a different context (introducing 
the first measurements of the gravitational
redshift of the White Dwarf companion of Sirius$^{10}$):
\begin{quote}
I have said that the observation was exceedingly difficult.  However
experienced the observer, I do not think we ought to put implicit
trust in a result which strains his skill to the utmost until it
has been verified by others working independently.  Therefore you should
for the present make the usual reservations in accepting these
conclusions. 
\end{quote}

For the next part of this section we turn from observation to theory,
in particular the construction and interpretation of calculations.

The mathematics laid out by Laplace, explained by Bowditch and referred to
by Herschel contain no errors.  That is, the manipulation of formulae
{\it once the problem had been set up} was correct.  But it is quite
possible to do the mathematical manipulations correctly and still
come up with the wrong answer.  In this episode we have examples
of three ways of
doing that: first, by ignoring something that could be important
(the motion of revolution of the ring); second, by making a mistake
in formulating the physical situation (calculating static stability for
an object not in equilibrium).

The third way is more subtle, lying in the transition from words to
mathematics: what do we mean by `stable?' Here,
there appears to have been a confusion of static with dynamic
stability, or perhaps a lack of recognition that there is a difference.
Herschel's use of the term {\it unstable equilibrium} and Laplace's
treatment of the symmetrical ring are unexeptionable---unless things are
in motion.  An object in what we might call a stable orbit around a planet, 
for instance, is {\it not}
in equilibrium; there are unbalanced forces on it and it is accelerating
all the time; so one must come up with a new definition in order to
deal with dynamic stability.

As noted above, Maxwell made the distinction and formulated a precise definition
of dynamic stability: a system is stable if a small disturbance gives a
small periodic deviation from the original motion.  There are other possible
definitions, for example this from the textbook of Thomson and Tait$^{11}$:
\begin{quote}
The actual motion of a system, from any particular configuration, is
said to be {\it stable} if every possible infinitely small conservative
disturbance of its motion through that configuration may be compounded
of conservative disturbances, any one of which would give rise to an
alteration of motion which would bring the system again to some
configuration belonging to the undisturbed path, in a finite time, and
without more than an infinitely small digression.  If this condition is
not fulfilled, the motion is said to be {\it unstable}.
\end{quote}
This is a rather more restrictive definition than Maxwell's: it requires
the disturbances to be conservative and a return, somewhere, to the original
path after a finite time.  One could imagine situations satisfying Maxwell's
definition and not Thomson and Tait's.  Whittaker$^{12}$ gives a choice of
definitions, one concerning deviations from a particular kind of
motion and attributed to Klein and Sommerfeld:
\begin{quote}
\ldots {\it steady motion} \ldots is defined to be a motion in which the
non-ignorable coordinates of the system have constant values, while
the velocities corresponding to the ignorable coordinates have also
constant values. 

\ldots

The steady motion is said to be {\it stable} if the vibratory motion
tends to a certain limiting form, namely the steady motion, when the
initial disturbance from steady motion tends to zero.  (p. 193, \S 83)
\end{quote}
Alternatively,
\begin{quote}
The word {\it stability} is often applied to characterise types of
motion in which the moving particle is confined to certain limited
regions \ldots (p.\ 417, \S 184)
\end{quote}
a definition which is noted as having been used by Hill, Bohlin and 
Darwin.  It is quite a different definition, mathematically, from
the previous ones, in that it allows deviations from a given path that
are not small.  Clearly there are situations satisfying this criterion
and not the others.

The general lesson to take away from this dictionary exercise is that
the translation from language (`is the ring stable?') to mathematics
may not be straightforward or single-valued, and unexpected subtleties
may appear.  This is the time to be especially clear and careful, or
you may wind up with mathematically elegant and sophisticated
calculations that don't, in fact, prove anything.

Finally, we move into the practical philosophy of science; that is,
how scientists (especially) expect the process of discovery to be
structured.  While its specific applicability to the episode of
Saturn's rings is speculative, I think the following passage from
Herschel's later book$^2$ is relevant:
\begin{quote}
Almost all the greatest discoveries in astronomy have resulted from
the consideration of what we have elsewhere termed RESIDUAL PHENOMENA,
of a quantitative or numerical kind, that is to say, of such
portions of the numerical or quantitative results of observations
as remain outstanding and unaccounted for after subducting and
allowing for all that would result from the strict application
of known principles. (p.\ 769, \S 856)
\end{quote}
Herschel thus expects to look for new discoveries in
the form of small quantities not previously discernable, perhaps by
using a newer, bigger telescope.  This is a 
reasonable expectation in any mature science (the big, obvious things have
already been accounted for) and arguably retains its utility as a general
rule.  The problem is, of course, that when one hunts for signals down
among the noise, one can mistake noise for signal.

A different point of practical scientific philosophy comes up in
the fact that, when two 
independent results agree, there is a very strong inclination
to conclude that they must both be true.  As expressed by Johannes
Kepler$^{13}$, `how easily the false disagrees with itself, and on the
other hand how reliably truth is consistent with truth' (Ch.\ I).
But sometimes the false agrees with itself.
{\em Sometimes errors reinforce each other}, as well as true results.
Kepler$^{13}$, again, was
familiar with this:
\begin{quote}
Indeed it is not the least important part of being shrewd to beware of
accidental associations of this kind, which, as the Sicilian siren
once detained seafarers with her singing, detain those engaged in
philosophy by the pleasure of their apparent beauty and their neatness
of fit \ldots so that they cannot attain the predetermined goal of
knowledge. (Ch.\ XIV)
\end{quote}

The phenomenon of reinforcing errors is the most important lesson of this
episode, the one to retain when all the details are gone.  I suspect
most scientists do not bear it sufficiently in mind, and most non-scientists
aren't aware of it at all.  But it can confidently be expected to appear
whenever one is searching for patterns amidst the noise on the edge of
current research.

\vspace{1.0cm}
\begin{center}
{\it Acknowledgements}
\end{center}
Much of the non-mathematical material in this article appears in or paraphrases
sections of the author's {\it Hindsight and Popular Astronomy}, \copyright
2011 World Scientific Publishing Inc.\ and used by their permission.  
Struve's article in {\it Astronomische Nachrichten} was kindly 
translated from the original German for me by Dr.\ Inga Schmoldt.
Much help, including locating Laplace's 1785 memoir on line, was
received from the librarian of the Institute of Astronomy of the
University of Cambridge, Mark Hurn; and the librarians of the
Cambridge University Library Rare Books room were unfailingly helpful.
\vspace{1.0cm}
\begin{center}
{\it References}
\end{center}

\noindent (1) Sir John F. W. Herschel, {\em Treatise on Astronomy}, 
New Edition (Longman,
Rees, Orme, Brown, Green \& Longman, and John Taylor, London), 1833 \\
\noindent (2) Sir John Herschel, {\em Outlines of Astronomy}, eleventh
edition, 1869; American edition (P. F. Collier \& Son, New York), 1902 \\
\noindent (3) A. F. O'D. Alexander, {\em The Planet Saturn: A History
of Observation, Theory and Discovery} 
(Faber and Faber, London), 1962, pp.\ 122-3 \\
\noindent (4) la Place, Pierre Simon marquis de, {\em Th\'{e}orie
des attractions des Sph\'{e}roides et de la figure des Plan\`{e}tes},
Histoire de l'Acad\'{e}mie Royale des Sciences, anne\'{e} M.DCCLXXXII,
pp.\ 113-196 of M\'{e}moires
(printed 1785); Biblioth\'{e}que Nationale de France, Gallica
Biblioth\'{e}que Num\'{e}rique (on-line facsimilie), starting on
\begin{verbatim}
http://gallica.bnf.fr/ark:/12148/pbt6k35819.image.f292.langFR
\end{verbatim} 
\noindent (5) la Place, Pierre Simon marquis de, {\em
M\'{e}canique C\'{e}leste}, volume I, 
translated with commentary by Nathaniel Bowditch
(Hilliard, Gray, Little and Wilkins, Boston), 1829 \\
\noindent (6)  \emph{Philosophical Magazine} (of London), July 1828 \\
\noindent (7) F.\ G.\ W.\ Struve, \emph{Astronomische Nachrichten}, 
{\bf 139}, 389, 1828 \\
\noindent (8) O.\ Struve, Sur les Dimensions des Anneaux de Saturne, in
{\em Recueil de M\'{e}moires Pr\'{e}sent\'{e}s \`{A} l'Acad\'{e}mie
Imp\'{e}riale des Sciences} (l'Acad\'{e}mie Imp\'{e}riale des
Sciences, Poulkowa), 1853, pp.\ 349-354 (Struve's
contribution is dated 15 November 1851) \\
\noindent (9) James Clerk Maxwell, On the Stability of Motion of Saturn's
Rings, in W.\ D. Niven ed., {\em The Scientific Papers of 
James Clerk Maxwell, Vol. I}
(Cambridge University Press, Cambridge), 1890, pp.\ 288-376 \\
\noindent (10) Sir A.\ S.\ Eddington, \emph{Stars and Atoms}, 
third impression (Yale University Press, New Haven and 
Oxford University Press, London), 1928, p.\ 53\\
\noindent (11) W. Thomson, Lord Kelvin and P. G. Tait, \emph{Treatise
on Natural Philosophy} (Cambridge University Press, Cambridge), 
1896, p.\ 416, \S 347 \\
\noindent (12) E. T. Whittaker, \emph{A Treatise on the Analytical
Dynamics of Particles \& Rigid Bodies} 4th ed. (Cambridge University
Press, Cambridge), 1947 \\
\noindent (13) J. Kepler, \emph{Mysterium Cosmographicum}, A. M. Duncan
trans. (Abaris Books, New York), 1981 \\
\end{document}